\documentclass[aps,twocolumn,preprintnumbers, amsmath, amssymb, floatfix]{revtex4}
\usepackage{dcolumn}
\usepackage{bm}

\def\be{\begin{equation}}
\def\ee{\end{equation}}
\def\beq{\begin{eqnarray}}
\def\eeq{\end{eqnarray}}

\begin{document}
\preprint{smw-ahw-10-L01}
\title{Explaining Michelson-Morley without Special Relativity}

\author{Sanjay M. Wagh}
\affiliation{Central India Research Institute, Post Box 606,
Laxminagar, Nagpur 440 022, India} \email{waghsm.ngp@gmail.com}

\author{Abhijit H Wagh}
\affiliation{Visiting Student Research Program, Central India
Research Institute, Post Box 606, Laxminagar, Nagpur 440 022,
India} \email{jetwagh@gmail.com}

\date{August 19, 2010}
\begin{abstract}
In this paper, we first discuss the concept of an emission wave.
In the history of science, this is the first time we have
discovered a new way in which (transverse) waves are realized in
nature. It can therefore be expected to lead to important changes
in the perspective about the nature of light or radiation. Then,
we point out that the null result of the Michelson-Morley
experiment is a natural and straightforward consequence of light
being an emission wave. Concepts of special relativity, of length
contraction and of time dilation, are not required for this
explanation, however.
\end{abstract}
\maketitle

\section{Introduction} \label{intro}

The particle and wave concepts are, as is well known, not mutually
exclusive always, but can be mutually supplementary. This is the
case when, as an example, particles undergo oscillatory or wavy
motion as a result of appropriate forces acting on them, and the
wave can be looked upon as some disturbance propagating within a
medium made up of those particles.

But, this supplementary nature of the particle and wave concepts
does not validate itself for, in particular, radiation. No forces
can be imagined to act on radiation (propagating in the vacuum),
and the quanta of radiation cannot be imagined to undergo
oscillatory motions.

With Louis de Broglie's seminal hypothesis that $\lambda=h/p$ for
a body of momentum $p=mv$, the lack of this supplementary nature
of the particle and wave concepts got extended to even material
bodies of mass $m$.

In describing the {\em Fresh Fields\/} explored during the years
1926-27, the era of the development of the quantum theory,
Heisenberg recalled \cite{heisenberg-01} that: {\em Bohr was
trying to allow for the simultaneous existence of both particle
and wave concepts, holding that, though the two were mutually
exclusive, both together were needed for a complete description of
atomic processes.}

But, the question of why any ``wave'' associates with a quantum
has not been addressed by this theory. In fact, it cannot be
addressed within this framework, this being its assumption.

The ``mutually exclusive'' character of these two concepts,
nevertheless, underlies the modern \cite{qmech} quantum theory,
which further turned out to be an intrinsically probabilistic
framework.

Bohr considered the particle and wave concepts as being
``complementary'' to each other. He then developed the principle
of the complementarity of concepts as the Copenhagen
Interpretation of the Quantum Theory.

\section{Concept of Emission Wave} \label{emit-wave}

But, any ``other type of'' supplementary nature of the particle
and wave concepts, than that to be found within the considerations
of Section \S\ref{intro}, was not explored in the past.

Nevertheless, the following is the other type of the supplementary
nature of the particle and wave concepts. {\bf \em This is,
previously unrealized, new way in which wave phenomena happen.} In
the history of science, this is the first time we have discovered
a new way in which (transverse) waves occur in nature. The
associated mental picture is as follows.

Consider a hosepipe emitting water through its nozzle. Let the
molecule of water move with some speed along a straight line after
its emission at the nozzle and let its speed not change, unless
that molecule happens to collide with another object along its
rectilinear path of motion.

Change in the location of the nozzle also changes the flux of
water passing location directly facing it. Any oscillatory motion
of the nozzle then produces oscillatory change in the flux of
water passing any location. The oscillatory changes in the
location of the nozzle are ``the cause'' for the oscillatory
changes in the jet of water.

No molecule is undergoing oscillatory motion. The oscillatory
spatiotemporal changes in their flux exist, still. The particle
and wave concepts are mutually supplementary, here.

We will, henceforth, refer to this wave as an emission wave.
Molecule of water can, clearly, be replaced by ``the quantum'' of
any kind, even the massless one. Thus, the emission waves of
quanta can get produced by their emitter, undergoing sort of an
oscillatory motion.

The massless quanta can be assumed to be momentum-less, that is,
as parcels of only energy. When emitted by a body, the direction
of the emitted quantum can be assumed to be spherically
symmetrically distributed around that instantaneous location of
its emitter.

If the wave phenomena of radiation arise indeed due to aspects of
emission of quanta, then such an origin must be consistent with
Planck's law. In other words, the aforementioned emission origin
for the wave of radiation quanta or the concept of an emission
wave needs to be consistent with the laws of the black body
radiation.

As was shown in \cite{eofwoq}, this is indeed the situation. In
summary, \cite{eofwoq} showed the following.

From only statistical considerations, the average number of the
quanta of radiation, of energy $\epsilon$, in equilibrium within a
cavity is given by \[ \left<n\right> = \frac{1}{
e^{\epsilon/kT}-1} \] On the other hand, the number of standing
wave modes within the frequency range $\nu$ to $\nu+d\nu$, at
frequency $\nu$ and enclosed within a cubical cavity of sides
$\ell$ is
\[ f(\nu)d\nu = \frac{8\pi\nu^2\ell^3}{c^3}d\nu \]
The spectral energy density of quanta contained within these modes
is then: \[ \epsilon\left< n\right> \, \frac{f(\nu)d\nu} {\ell^3}
= \frac{8\pi\epsilon\nu^2}{c^3} \frac{d\nu}{e^{\epsilon/kT}-1} \]

This last expression reduces to Planck's formula for the spectral
energy density of the black body radiation only when we assume
$\epsilon=h\nu$.

{\em However, the energy $\epsilon$ of a light-quantum is not
related to the frequency $\nu$ of the wave of Light, for nowhere
is this relation implied by this mechanism.} Nevertheless, both
the particle and wave pictures are implied by this mechanism,
albeit in the roles supplementary to each other. Bohr's point of
view that the particle and wave pictures, both together, are
needed for the complete description of atomic processes is clearly
justified within this concept of an emission wave.

\section{Propagation of Emission Wave}
Of very specific interest now is the fact that the emission wave
does not require any ``pre-existing medium'' for its propagation.
This follows from the following considerations.

{\bf \em An emission wave propagates away from the nozzle, its
source, only as the molecules of water propagate away from it. The
wave gets ``created'' along with the emission of molecules at the
nozzle and ``propagates'' along with them. Any prior existence of
the ``medium of water'' is not necessary for the propagation of
the emission wave. }

This above is applicable to the emission wave of quanta. In
particular, for the radiation quanta, this holds. The quantum of
radiation can also be momentum-less and mass-less, as these
properties are unrelated to the mechanism of the creation of the
emission wave of such quanta.

\section{Michelson-Morley type Experiments}
During the beginning of the twentieth century, certain experiments
attempted detection of \ae ther, the medium of propagation of
light. The famous are the experiments of Fizeau, and of
Michelson-Morley, which had relied on the phenomenon of the
interference of light.

The reasoning underlying these experiments is that of the
luminiferous \ae ther being carried along with them by moving
bodies.

If light were wave propagating within an existing medium of \ae
ther, then its speed in the direction of motion of the \ae ther
would be different than its speed opposite to the direction of
motion of the luminiferous \ae ther.

Then, if (monochromatic) light propagating in the direction of and
that propagating opposite to the direction of motion of the \ae
ther were made to interfere in an interferometer, we would obtain
the interference fringes due to the path difference induced by
difference in the speed of propagation of light in these two
situations.

If a dark fringe were to be prearranged in the interference
apparatus, then we should detect shift in the fringe as the
apparatus moves about in the space. Detecting this shift had been
the aim of the experiments mentioned before.

As is well known, all these experiments did not detect any \ae
ther. The absence of the fringe shift in such experiments is then
explained under the Lorentz transformations of space-time
coordinates. This involves time-dilation and length contraction,
the latter was first pointed out by Fitzgerald as a ``natural
consequence'' of propagation of a body through the \ae ther.

\section{Explaining Michelson-Morley type experiments without Special
Relativity}

Now, for these experiments, we consider light as a monochromatic
emission wave. The source of the light quanta is assumed to be
oscillating, in this case, with a single frequency $\nu$, and the
quanta of light to be propagating with the speed of light (in
vacuum), all. The frequency of the emission wave of light is then
also $\nu$.

Under the assumption that the quanta of light are mass-less, the
speed of propagation of their emission wave is the speed of the
quanta themselves. The velocity of mass-less  and momentum-less
quanta cannot be altered as per the concepts of Galileo and
Newton. This velocity is the {\em same\/} for all the observers.

The emission wave of radiation propagates from the source to its
reflector. On its reflection, the emission wave returns to the
origin, and is made to interfere with the main wave. The speed of
the emission wave is always the same, irrespective of the
direction of its propagation, in vacuum and in any material like
water in the case of Fizeau's celebrated experiment.

As no medium of propagation is involved, if the destructive
interference is pre-arranged within the interferometer, then the
dark line will not shift even if the apparatus, {\em ie}, the
interferometer, were to move about in the space.

This is an explanation of the Michelson-Morley experiment.
Clearly, we do not require the length contraction or time dilation
(and the framework of the special relativity) to explain the
results of the Michelson-Morley kind of experiments attempting to
detect the medium, luminiferous \ae ther, of the propagation of
light.


In 1927, Dirac stated \cite{dirac-01} that {\em the light quantum
has the peculiarity that it apparently ceases to exist when it is
in $\cdots$ the zero state in which its momentum, and therefore
its energy, is zero. When a light quantum is absorbed it can be
considered to jump into the zero state, and when one is emitted it
can be considered to jump from the zero state to one in which it
is in physical evidence, so that it appears to have been created.}

This picture is then needed along with the emission view for the
wave of light, and the massless, momentum-less, nature for the
energy quanta of light. The light beam reflected by the reflector
is then produced as a result of the complete absorption of the
energy quanta incident on it, and the total re-emission of the
energy quanta constituting the reflected beam of light.

The mathematical framework underlying such processes is
\cite{smw-utr, smw-let-0601, smw-sars, smw-ppt, lir-gr, mleq} that
of the Universal Relativity. The details of this framework are
however beyond the natural scope of the present discussion.

\acknowledgments We are grateful to Dilip Deshpande and Sandeep
Deshpande for helpful discussions.

This article is dedicated to fond memories of (Late) Professor P C
Vaidya and (Late) Professor Geoffrey Burbidge, we missing the
opportunity of hosting the latter during an IAGRG Meeting held at
Nagpur in 2001.

\end{document}